\documentclass[aps,prb,reprint,twocolumn,superscriptaddress,floatfix,nofootinbib]{revtex4-2}
\usepackage{epsfig,amsmath,amssymb,color,comment,physics, dsfont, bbold}
\usepackage[makeroom]{cancel}
\usepackage[caption=false]{subfig}
\usepackage{mathrsfs}
\usepackage[countmax]{subfloat}
\usepackage[normalem]{ulem}
\usepackage[english]{babel}
\usepackage{dsfont}
\usepackage{float}
\usepackage{color}
\usepackage[dvipsnames]{xcolor}
\usepackage{tikz}

\usetikzlibrary{quantikz}

\usepackage[bookmarks=true,colorlinks,linkcolor=blue,urlcolor=NavyBlue,citecolor=RoyalBlue]{hyperref}

\begin{document}

\title{Non-stabilizerness in kinetically-constrained Rydberg atom arrays}
\author{Ryan Smith}
\affiliation{School of Physics and Astronomy, University of Leeds, Leeds LS2 9JT, UK}

\author{Zlatko Papi\'c}
\affiliation{School of Physics and Astronomy, University of Leeds, Leeds LS2 9JT, UK}

\author{Andrew Hallam}
\affiliation{School of Physics and Astronomy, University of Leeds, Leeds LS2 9JT, UK}

\date{\today}
\begin{abstract}
    Non-stabilizer states are a fundamental resource for universal quantum computation. However, despite broad significance in quantum computing, the emergence of ``many-body'' non-stabilizerness in interacting quantum systems remains poorly understood due to its analytical intractability. 
    Here we show  that Rydberg atom arrays provide a natural reservoir of non-stabilizerness that extends beyond single qubits and arises from quantum correlations generated by the Rydberg blockade. We demonstrate that this non-stabilizerness can be experimentally accessed using two complementary methods, either performing quench dynamics or via adiabatic ground state preparation. Using the analytical framework based on matrix product states, we explain the origin of Rydberg non-stabilizerness via a quantum circuit decomposition of the wave function.         
\end{abstract}
\maketitle

\section{Introduction}\label{Section:Intro}

Concepts from quantum information theory have become indispensable tools for understanding many-body quantum systems. For example, quantum entanglement is now central to the understanding of topological order~\cite{Kiteav2006,Levin2006,Li2008,Pollmann2012} and the non-equilibrium dynamics of interacting quantum systems~\cite{Calabrese_2005,Bardarson2012,Serbyn2013,lukin2019} (see Ref.~\cite{Laflorencie2016} for a review). Moreover, low-energy eigenstates of quantum Hamiltonians typically have a limited amount of entanglement, making them amenable to variational ansatze such as tensor networks and the density-matrix renormalization group~\cite{WhiteDMRG,SCHOLLWOCK201196,CiracRMP}. 

Entanglement, however, is not the only resource required for large-scale, fault tolerant, quantum computation. The implementation of a universal set of gates is a major challenge -- typically only the Clifford group of multi-qubit Pauli gates is feasible, a set of gates that can be efficiently simulated classically \cite{gottesman1998,Eastin2009}. One approach to universal quantum computation is to inject non-stabilizer or ``magic'' states into the circuit \cite{knill2004,Bravyi2005,Campbell2010}. From a practical standpoint, this raises the question how such special states can be conveniently generated. While there has been much progress in understanding the non-stabilizerness of \emph{few-qubit} systems~\cite{Bravyi2005,Howard2017}, analogous property in \emph{many-qubit} Hamiltonian and circuit systems is a subject of active investigation~\cite{ShiyuTgate2020,goto2021chaos,Liu2022,Leone2022,Haug2023,Tarabunga2023,Turkeshi2023,bejan2023dynamical,niroula2024phase,Tarabunga2024,haug2024probing}. Consequently, many basic questions remain open, e.g., whether non-stabilizerness can play a similar role to entanglement in characterizing the properties of many-body systems.

 Unfortunately, quantifying the non-stabilizerness of generic wave functions is very costly. In this work we focus on the Stabilizer R\'enyi Entropy (SRE)~\cite{Leone2022,Oliviero2022}, which has recently been proposed as a measure of non-stabilizerness for many-qubit wave functions (we note there are related local measures such as ``robustness of magic"~\cite{Heinrich2019robustnessofmagic,Howard2017,Sarkar_2020} and Mana entropies for qudit systems~\cite{tarabunga2024critical,frau2024nonstabilizerness}). The SRE, while still exponentially hard to evaluate in general, can be approximated via Monte Carlo methods~\cite{Tarabunga2023,lami2023quantum,Liu2024}. Morever, for a class of matrix product states (MPS) with a low bond dimension, the SRE can be expressed in closed form~\cite{Haug2023,Haug2023stabilizerentropies,Banuls2024}. Nevertheless, it remains unclear what information about the wave function is contained in its SRE. Studies of quantum spin chains and related models have empirically found that many-body ground states can exhibit varying levels of non-stabilizerness but generally do not approach the upper bound of the SRE, even at quantum critical points~\cite{Tarabunga2023}. Moreover, a microscopic explanation of the origin of the SRE for a given state is lacking. Thus, it is important to identify analytically tractable models where enhancement of SRE compared to a single qubit can be analytically understood.

 In this work we study non-stabilizerness in the PXP model -- an effective model of one-dimensional (1D) Rydberg atom quantum simulators~\cite{Bernien2017,Turner2018,TurnerAPS}. 
 This model has the advantage that both its ground state and certain kinds of far-from-equilibrium quench dynamics can be accurately described using a manifold of low-bond dimension MPS states~\cite{Ho2019,Michailidis2020}. While the ground states of gapped systems in 1D are well-known to be captured by finite bond dimension MPS, the ability to describe highly excited quench dynamics in the same way is due to the PXP model hosting a set of non-thermalizing eigenstates known as quantum many-body-scars (QMBSs)~\cite{Serbyn2021Review, Papic2022Entanglement, Moudgalya2022Review, Chandran2023Review}.   We analytically calculate the SRE across the MPS manifold associated with QMBS  dynamics of the PXP model, finding that the manifold hosts regions of large non-stabilizerness. 
 The non-stabilizerness can be understood microscopically from the corresponding unitary circuit that generates the states in the manifold. Finally, we demonstrate that the MPS manifold is relevant for the physics of the full 1D Rydberg model realized in experiments, and we propose two protocols that can be used to access the non-stabilizer states, either by performing the global quench or by adiabatic ground state preparation. 

 The remainder of this paper is organized as follows. In Sec.~\ref{sec: SRE} we review the concept of stabilizer R\'enyi entropy for MPS states -- the main quantity of interest in this work. In Sec.~\ref{Sec:PXP} we introduce the PXP model of Rydberg atom arrays and the MPS ansatz that describes its non-equilibrium behavior under quench dynamics from certain initial states. Focusing on this MPS variational manifold, we then analyze the non-stabilizerness that can be generated under time evolution and we identify its microscopic origin. Two experimental protocols for observing the non-stabilizerness in Rydberg atom arrays are discussed in Sec.~\ref{Sec: Experiment}. Our conclusions are presented in Sec.~\ref{Sec:Conclusions}, while Appendices contain further technical details of the MPS calculations, a discussion of other measures of non-stabilizerness, and a study of non-stabilizerness of the eigenstates of the PXP model. 

\section{Stabilizer R\'enyi entropy of MPS states}\label{sec: SRE}

 For a pure state $\ket{\psi}$ of $N$ spin-1/2 particles, a useful measure of non-stabilizerness is the stabilizer Renyi entropy (SRE) of order $n$~\cite{Leone2022}:
\begin{equation}\label{eq:SRE}
M^{(n)}(\ket{\psi})=(1-n)^{-1}\ln{\sum_{P\in\mathcal{P}_N}\frac{\bra{\psi}P\ket{\psi}^{2n}}{2^N}},
\end{equation}
where  $\mathcal{P}_N$ denotes the set of all $N$-strings of Pauli matrices $\{\sigma^\alpha\}=\{\mathbb{I},\sigma^x,\sigma^y,\sigma^z\}$. 
The SRE is zero iff $\ket{\psi}$ is a stabilizer, it is invariant under Clifford unitaries and additive under tensor product~\cite{Leone2022}.

The cost of directly evaluating Eq.~\eqref{eq:SRE} scales as $4^N$, which rapidly becomes intractable. Instead, the SRE can be approximated by Monte Carlo sampling over Pauli strings \cite{Tarabunga2023,lami2023quantum}, which scales more favorably but may require many samples to obtain accurate statistics. Finally, the SRE can be calculated directly using MPS techniques, naively scaling like $\chi^{6n}$ in the bond-dimension $\chi$ of the MPS~\cite{Haug2023}. We utilize the latter approach in this work as we focus on low-$\chi$ MPS states for which we can obtain analytical insight into the SRE.

\subsection{Replica MPS method for evaluating SRE}

We consider a translation-invariant MPS state $\ket{\psi(A)}$, defined on an infinite lattice with $(d=2)$-dimensional local Hilbert space, 
\begin{eqnarray}
 \ket{\psi(A)} = \sum_{\{\sigma_{j}\}} \mathrm{tr}(\cdots A^{\sigma_{j-1}}A^{\sigma_{j}}A^{\sigma_{j+1}} \cdots) \ket{ \{ \sigma_j \}},   
\end{eqnarray}
where $A^{\sigma_j}$ is a set of $d$ matrices of dimensions $\chi \times \chi$, with $\chi$ being the MPS bond-dimension. To calculate the SRE of $\ket{\psi(A)}$, we employ the replica trick method from Ref.~\cite{Haug2023} that we briefly review in this section. 

To calculate the $n$th order SRE, we begin by creating a $2n$-fold replica of the state $\ket{\phi^{(n)}}=\ket{\psi}^{\otimes 2n}$ with physical dimension $d^{\prime}=d^{2n}$ and bond dimension $\chi^{\prime}=\chi^{2n}$. Below we will mainly be interested in the simplest $n=2$ case.
Let us define tensors
\begin{eqnarray}
    B_{i,j}^{\sigma^{s}} = (A^\sigma_{ij})^{\otimes2n}, \quad 
    \Lambda_j^{(n)} = \frac{1}{2} \sum^{3}_{\alpha=0}(\sigma^\alpha_j)^{\otimes 2n},
\end{eqnarray}
where $B_{i,j}^{\sigma^{s}}$ is a $2n$-fold copy of $A$ and  $\Lambda^{(n)}_{j}$ acts over every $2n$ replica of $\ket{\psi}$ on a physical site $j$. With these tensors,  the expectation value over all the $N$-qubit Pauli strings can be seen as a single expectation value:
\begin{equation}\label{eq:finitemagic}
\sum_{P\in\mathcal{P}_N}\frac{\bra{\psi}P\ket{\psi}^{2n}}{2^N}=
\bra{\phi^{(n)}} \Lambda^{(n)}_1 \otimes ... \otimes \Lambda^{(n)}_N \ket{\phi^{(n)}},
\end{equation}
which we will denote as $\bra{\phi^{(n)}}\boldsymbol{\Lambda}^{(n)}\ket{\phi^{(n)}}$. Thus, the many-body SRE is:
\begin{equation}
    M^{(n)}(\ket{\psi}) = \frac{1}{1-n} \ln\bra{\phi^{(n)}}\boldsymbol{\Lambda}^{(n)}\ket{\phi^{(n)}},
\end{equation}
and the SRE density can be defined as 
\begin{equation}\label{eq:SREdens}
 m^{(n)}=M^{(n)}/N.   
\end{equation}
Below we will always use the SRE density $m^{(n)}$ as it is an intensive quantity that can also be defined for a state on an infinite lattice, discussed next.

The SRE in the thermodynamic limit is obtained from the modified $\chi^{4n}{\times}\chi^{4n}$ transfer matrix:
\begin{equation}\label{eq:transfermatrix}
\tau_{(ik),(jl)}=\sum_{s,s^\prime} B^{\sigma^s}_{i,j}(\Lambda^{(n)}_1 \otimes \cdots \otimes \Lambda^{(n)}_N)^{\sigma^s,\sigma^{s^\prime}}  \bar{B}^{\sigma^{s^\prime}}_{k,l}.
\end{equation}
Denoting the dominant eigenvalue of $\tau$ as $\lambda_{0}^{(n)}$, the SRE density in the thermodynamic limit is given by 
\begin{eqnarray}\label{eq:m2}
 m^{(n)}(\ket{\psi(A)}) = \frac{1}{1-n} \ln{\lambda^{(n)}_0}.
\end{eqnarray}
As defined here, the SRE is upper-bounded by 
\begin{equation}\label{eq:SREbound}
 m^{(n)}\leq (1/N)\ln \mathcal{D},   
\end{equation}
where $\mathcal{D}$ is the Hilbert space dimension~\cite{Leone2022}. 

Performing the SRE calculation directly is computationally demanding for $\chi \gtrsim 8$. Thus, for larger bond dimensions we make use of a reformulated version of the replica trick via conversion to the Pauli basis~\cite{Banuls2024}. This allows for truncation of the bond dimension during the calculation and the key steps are outlined in Appendix~\ref{App:pauli basis}.

\subsection{Global vs local non-stabilizerness}\label{subsec:global vs local}

The SRE, Eq.~(\ref{eq:SRE}),  is intrinsically a global property of the wave function as it depends upon $N$-site products of Pauli matrices, and the latter can be extended throughout the entire system. In contrast, physical observables are typically spread over a few physical sites and therefore can be measured on a finite subsystem described by the reduced density matrix. We will now address the extent to which the non-stabilizerness of the MPS ansatz is stored in local or global degrees of freedom. 

The SRE is a basis dependent quantity, hence it can vary even for uncorrelated product states.  For this reason, it is common to consider ``long-range'' non-stabilizerness, defined as the minimal non-stabilizerness after performing arbitrary local rotations on every site of the system. As an example, for the MPS ansatz that will be used in the PXP model in Sec.~\ref{Sec:PXP} below, the only relevant rotation that can reduce the SRE is the $y$-axis rotation for spins on the even and odd sublattice:
\begin{eqnarray}
 \ket{\psi}\rightarrow \ket{\tilde{\psi}(\gamma_o,\gamma_e)}=\bigotimes_j(e^{i\gamma_o\sigma_{2j-1}^y}\otimes e^{i\gamma_e\sigma_{2j}^y})\ket{\psi}.  \;\;
\end{eqnarray}
The long-range SRE is then defined by minimizing over the local basis rotations:
\begin{equation}\label{eq:longrangeSRE}
m^{(2)}_L=\min_{\{\gamma_o,\gamma_e\}}m^{(2)}\Big(\ket{\tilde{\psi}(\gamma_o,\gamma_e)}\Big).
\end{equation}

Another way to distinguish local from global non-stabilizerness is via the non-stabilizerness of a two-site reduced density matrix $\rho_2$. Since $\rho_2$ is generally a mixed state, the calculation of the SRE density changes from the definition given in Eq.~\eqref{eq:SRE} to~\cite{Leone2022}:
\begin{equation} \label{Eq:mixed_state_SRE}
    m^{(2)}(\rho_2)=-\frac{1}{2}\ln\left(\frac{\sum_{P \in \mathcal{P}_2}|\mathrm{tr}(\rho_2 P)|^4}{\sum_{P \in \mathcal{P}_2}|\mathrm{tr}(\rho_2 P)|^2}\right).
\end{equation}
Therefore, the mixed state magic is computed from the expectation values of the 16 two-qubit Pauli operators in the state $\rho_2$. For simplicity, we will restrict to two-site reduced density matrices and denote $\rho_2$ simply as $\rho$. 

We note that SREs are not strong magic monotones, i.e., they do increase under non-deterministic stabilizer protocols \cite{Leone2024Monotones}. However, they are widely used as they are relatively easy to compute for large systems. In Appendix ~\ref{Sec:ROM} we consider the ``Robustness of Magic" (RoM) of a two-qubit reduced density matrix $\rho_2$ as a method of quantifying local magic. Unlike SREs, the RoM is a strong magic monotone but is typically reserved for few qubit systems.

\section{Non-stabilizerness in the PXP Model}\label{Sec:PXP}

In the remainder of this paper, we focus on the physical manifestations of non-stabilizerness in Rydberg atom arrays~\cite{Browaeys_review}. The latter are described by the kinetically constrained 1D spin-$1/2$ PXP model~\cite{FendleySachdev,LesanovskyKatsura,Turner2018}:
\begin{equation}\label{Eq:PXP model}
    H_\mathrm{PXP} = \frac{\Omega}{2}\sum_{j=1}^N P_{j-1}\sigma_{j}^x P_{j+1},
\end{equation}
 where $\Omega=1$ is the Rabi frequency describing the flipping of each atom between its ground state $|0\rangle$ and the excited Rydberg state $|1\rangle$, $N$ is the total number of atoms, and $P_j = (1-\sigma_j^z)/2$ is the projector on the $\ket{0}$ state. With open boundary conditions we set $P_{0} = P_{N{+}1} = \mathbb{1}$. The PXP model describes the low-energy physics in the Rydberg blockade regime~\cite{Browaeys_review}, a phenomenon where neighboring excitations of the atoms, such as $\ket{...11...}$, are energetically forbidden. The blockade is imposed globally with the projector $\mathcal{P}=\bigotimes_j (\mathbb{1}-\ket{11}\bra{11}_{j,j+1})$ or, equivalently, in the local form using the $P_j$ operators as in Eq.~(\ref{Eq:PXP model}). 

\begin{figure*}
    \centering
    \includegraphics[width=0.99\linewidth]{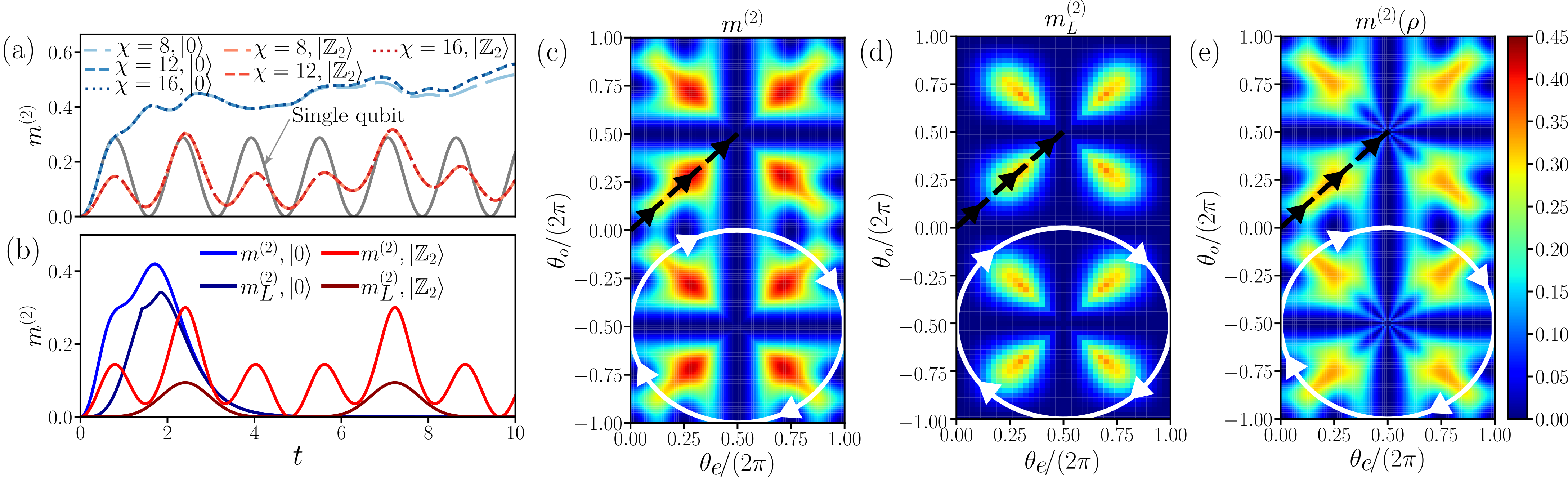}
    \caption{ (a)-(b): The dynamics of $m^{(2)}$ following the quench from the $\ket{0}$ and $\ket{\mathbb{Z}_2}$ initial states in the PXP model.
    Panel (a) shows full quantum dynamics obtained numerically using TDVP for  $N=51$ spins and different bond dimensions $\chi$ indicated in the legend. Single-qubit precession is shown in gray for comparison. Panel (b) shows the dynamics projected to $\mathcal{M}$ with $\chi=2$, obtained analytically by integrating the equations of motions~(\ref{eq:TDVPanalytic}). For comparison, we also include the long-range non-stabilizerness, $m^{(2)}_L$, in Eq.~(\ref{eq:longrangeSRE}).   (c)-(e): Phase diagram of non-stabilizerness across the MPS manifold $\mathcal{M}$, plotted as a function of $\theta_e$ and $\theta_o$ angles. 
    The three panels correspond, respectively, to $m^{(2)}$, its long-range version $m^{(2)}_L$ in Eq.~(\ref{eq:longrangeSRE}), and the two-site mixed state non-stabilizerness $m^{(2)}(\rho)$ in Eq.~(\ref{Eq:mixed_state_SRE}). The trajectories traversed by the $\ket{\mathbb{Z}_2}$  and $\ket{0}$ states are shown by solid white and black dashed lines, respectively. 
    }
    \label{fig:PXP Magic}
\end{figure*}

\subsection{PXP dynamics and MPS ansatz}

When the PXP model is quenched from  typical initial states such as $\ket{0}{=}\ket{0000 \cdots}$, rapid thermalization is observed~\cite{Bernien2017}, consistent with the system being chaotic~\cite{Turner2018}. Nevertheless, the PXP model also hosts a small number of non-thermal eigenstates which are evenly distributed in energy and possess anomalously low entanglement entropy~\cite{Turner2018,TurnerAPS,Lin2019,Iadecola2019,Khemani2018,Choi2018,Bull2020,Omiya2022}. These QMBS eigenstates have a high overlap with the  $\ket{\mathbb{Z}_2}{=}\ket{0101\cdots}$ state, leading to a suppressed growth of entanglement entropy and periodic revivals in local observables when the system is quenched from that initial state~\cite{Bernien2017}.

The short-time dynamics of the PXP model can be understood semiclassically using an MPS ansatz with $\chi=2$ with the help of time-dependent variational principle (TDVP)~\cite{Ho2019,Michailidis2020}. The ansatz is physically motivated by applying the Rydberg blockade projector $\mathcal{P}$ to a product of spin coherent states, 
\begin{eqnarray}
\nonumber  |\psi(\boldsymbol{\theta},\boldsymbol{\phi})\rangle = \mathcal{P}\bigotimes_{j} \Big( \cos(\theta_{j}/2)|0\rangle_{j}-ie^{i\phi_j}\sin(\theta_{j}/2) |1\rangle_j \Big),  \\
\end{eqnarray}
where $\theta_j$, $\phi_j$ are the Bloch sphere angles of $j$th spin. 
The resulting state $\ket{\psi(\boldsymbol{\theta},\boldsymbol{\varphi})}$ can be equivalently expressed as a $\chi=2$ MPS~\cite{Ho2019}:
\begin{equation}\label{Eq:PXP MPS}
    A^{0}(\theta_j,\phi_j) = \begin{pmatrix}
        \text{cos}(\theta_j/2)& 0 \\
        \text{sin}(\theta_j/2) & 0 
    \end{pmatrix}, \; \; A^{1}(\theta_j,\phi_j) = \begin{pmatrix}
        0 & -ie^{i\phi_{j}} \\
        0 & 0 
    \end{pmatrix},
\end{equation}
which defines a continuous manifold  $\mathcal{M}=\mathrm{span}\{ \ket{\psi(A)} | \forall \theta_j, \phi_j \}$. The manifold $\mathcal{M}$, by definition, includes the states $|0\rangle$ and $|\mathbb{Z}_2\rangle$, which are representatives of generic (thermalizing) and scarred (non-thermalizing) initial conditions.

We note that the Rydberg blockade projector increases the periodicity of $\mathcal{M}$ to $T=4\pi$~\cite{Ho2019}. As we are primarily interested in $\ket{0}$ and $\ket{\mathbb{Z}_2}$ initial states, we assume two-site periodicity of the angles, leaving the angles on even and odd sites, $(\theta_e,\phi_e)$ and $(\theta_o,\phi_o)$, as our only parameters. Furthermore, we can set $\phi_e = \phi_o = 0$ due to energy conservation~\cite{Ho2019}. Quantum dynamics can then be projected to $\mathcal{M}$ using TDVP. This projection ``distorts'' the Schr\"odinger equation, resulting in a system of two non-linear differential equations for $\theta_e$ and $\theta_o$:
\begin{equation}\label{eq:TDVPanalytic}
    \begin{aligned}
        &\dot{\theta_o}=\cos(\frac{\theta_e}{2}) + \sin(\frac{\theta_o}{2})\tan(\frac{\theta_e}{2})\cos^2\left(\frac{\theta_o}{2}\right), \\
        &\dot{\theta_e}=\cos(\frac{\theta_o}{2}) + \sin(\frac{\theta_e}{2})\tan(\frac{\theta_o}{2})\cos^2\left(\frac{\theta_e}{2}\right),
    \end{aligned}
\end{equation}
see Ref.~\cite{Ho2019} for a derivation. This elegant description of the dynamics in terms of low-$\chi$ MPS states will allow us to gain analytical insight into the out-of-equilibrium behavior of non-stabilizerness in the PXP model, Eq.~(\ref{Eq:PXP model}).

\subsection{Non-stabilizerness of MPS ansatz}\label{Sec:PXP Magic}

The dynamics of SRE density $m^{(2)}$, Eq.~(\ref{eq:m2}), when the PXP model is quenched from $\ket{0}$ and $\ket{\mathbb{Z}_2}$ initial states is presented in Fig.\ref{fig:PXP Magic}(a).  We obtain the time-evolved state in MPS representation for various bond dimensions using a numerical implementation of TDVP~\cite{Haegeman} and then evaluate its SRE according to Eq.~(\ref{eq:finitemagic}). For the scarred $\ket{\mathbb{Z}_2}$ state, we see a complex pattern in the dynamics of SRE, with a single large peak in between two smaller peaks. This data is well-converged already with small $\chi$. The complex dynamics of SRE should be compared with the independent spin precession generated by $H{=}\sum_j \sigma^x_j/2$, which is also shown in Fig.\ref{fig:PXP Magic}(a). By contrast, the thermalizing $\ket{0}$ state displays a rapid increase in SRE, exceeding the values reached by the $\ket{\mathbb{Z}_2}$ state. Following this initial increase, the SRE of $|0\rangle$ initial state remains relatively stable, despite the continuously increasing entanglement entropy found in Ref.~\cite{Turner2018}. 

The dynamics projected into $\mathcal{M}$ lead to perfectly periodic evolution of the SRE when starting in the $\ket{\mathbb{Z}_2}$ state, see Fig.~\ref{fig:PXP Magic}(b). Although the full dynamics in Fig.~\ref{fig:PXP Magic}(a) is not exactly periodic, the TDVP representation of $m^{(2)}$ within $\mathcal{M}$ still shows excellent agreement. In particular, the distinctive pattern of a large peak surrounded by two smaller peaks in $m^{(2)}$ is fully reproduced within $\mathcal{M}$, and we will provide its explanation in Sec.~\ref{Sec:unitary} below. For the $\ket{0}$ initial state, however, the agreement between full dynamics and $\mathcal{M}$ is only good up to times $t \approx 2$, after which the TDVP clearly no longer captures the full SRE dynamics. This is expected due to the large leakage of the dynamics outside the manifold~\cite{Ho2019}. In fact, at late times the evolution of the $\ket{0}$ state becomes perpendicular to the ansatz and $m^{(2)}$ is stuck near zero. Nevertheless, comparing Fig.\ref{fig:PXP Magic}(a) and Fig.\ref{fig:PXP Magic}(b), we see that projection into $\mathcal{M}$ captures well the early-time enhancement of the SRE, which will be our main focus below.

Since non-stabilizerness is basis-dependent, \emph{a priori} it is not clear if the behavior seen in  Fig.\ref{fig:PXP Magic}(a) is truly a many-body phenomenon or if it can be removed through a local unitary rotation as discussed in Sec.~\ref{subsec:global vs local}. In Fig.~\ref{fig:PXP Magic}(b) the long-range SRE, $m_L^{(2)}$ in Eq.~(\ref{eq:longrangeSRE}), for the MPS ansatz trajectories of the $\ket{0}$ and $\ket{\mathbb{Z}_2}$ states are shown. For the $\ket{0}$ state, we see the long-range SRE is slightly smaller than the SRE but not significantly different, indicating that most of the SRE is `long-range' for this trajectory. For the $\ket{\mathbb{Z}_2}$ state, on the other hand, we see a more pronounced difference between $m^{(2)}$ and $m_L^{(2)}$. The distinctive three-peak behavior seen in the SRE disappears for the long-range SRE, with only a single residual peak, much reduced in magnitude. For this reason, we argue that the two smaller peaks of the SRE in the $\ket{\mathbb{Z}_2}$ case can be considered primarily local, i.e., similar in nature to the non-stabilizerness of a product state. The larger central peak, however, is a genuine many-body phenomenon, arising from the correlations due to the Rydberg blockade.

Figure~\ref{fig:PXP Magic}(c) shows the SRE across the manifold $\mathcal{M}$ plotted as a function of $\theta_e$ and $\theta_o$ angles. When either angle is zero, the MPS ansatz reduces to a simple product state, with one of the spins fixed to $\ket{0}$ and the other rotating in the $yz$-plane. Hence, the SRE reaches a maximum whenever it is furthest from the other spin being an $\sigma^{y}$ or $\sigma^{z}$ eigenstate, i.e., at $\theta_{o/e}=\pi/4, 3\pi/4,\ldots$. Cases $\theta_{o/e}=\pi$ are also easy to understand, since one of the sites in the unit cell is guaranteed to be occupied, hence the other must be unoccupied due to the Rydberg blockade. Since this corresponds to the $\ket{\mathbb{Z}_2}$ product state, the SRE is zero along these lines. The non-trivial feature of the diagram in Fig.\ref{fig:PXP Magic}(c) are the four arrow-like structures pointing towards the center of each quadrant. The SRE maxima lie at the centers of these arrows along the $\theta_e = \pm \theta_o$ line with a large value of $m^{(2)} \approx 0.42$. Note that the latter is considerably larger than the quantum Ising model at its critical point \cite{Tarabunga2023,Liu2024}. 

The TDVP trajectory followed by the $\ket{\mathbb{Z}_2}$ state in $\mathcal{M}$ is shown by a white solid line in Fig.~\ref{fig:PXP Magic}(c), moving periodically between high and low SRE regions. By contrast, the $\ket{0}$ state follows the black dashed line in Fig.~\ref{fig:PXP Magic}(c), moving diagonally from $(\theta_o,\theta_e)=(0,0)$ to $(\theta_o,\theta_e)=(\pi,\pi)$, passing through the state with maximal SRE before exiting the manifold. Recalling Fig.~\ref{fig:PXP Magic}(b), it is clear that the repeating pattern of three SRE peaks is due to the $\ket{\mathbb{Z}_2}$ trajectory cutting through the arrow-shaped regions in Fig.~\ref{fig:PXP Magic}(c). The two smaller peaks in the SRE are from the sides of the arrow shape,  suggesting they arise due to local non-stabilizerness, whereas the larger peak is a many-body effect. By contrast, the dynamics initialized in the $\ket{0}$ state directly flows towards the global non-stabilizerness maximum $m^{(2)} \approx 0.42$ within $\mathcal{M}$, close to the value in the full model in Fig. \ref{fig:PXP Magic}(a). Thus, high non-stabilizerness is created under PXP dynamics within $\mathcal{M}$, given that SRE is upper-bounded by $m^{(2)} \lesssim 0.48$ due to the PXP Hilbert space dimension growing as the Fibonacci number~\cite{TurnerAPS}. 

Figure~\ref{fig:PXP Magic}(d) shows the long-range SRE, Eq.~(\ref{eq:longrangeSRE}), across the entire MPS manifold. We see that the arrow-head structure vanishes for the long-range SRE, demonstrating that only the central areas of the arrow-heads are non-local. On the other hand,  the SRE of a two-site reduced density matrix, Eq.~(\ref{Eq:mixed_state_SRE}), plotted in Fig. \ref{fig:PXP Magic}(e) accurately recreates the structures obtained in the full SRE calculation. 
The two-site reduced density matrix is generally in good agreement with $m^{(2)}$ in Fig. \ref{fig:PXP Magic}(c), with some quanitative differences, e.g., the peak SRE for the former is at around $m^{(2)}(\rho)\approx0.32$, while for the latter it is considerably larger, $m^{(2)}\approx0.42$. 

In summary, comparing Fig.~\ref{fig:PXP Magic}(c)-(e) we conclude that a two-site reduced density matrix contains sufficient information to qualitatively reproduce the many-body non-stabilizerness that emerges after the quench from $|0\rangle$ initial state in the PXP model. While the two-site reduced density matrix, in principle, also captures the behavior of non-stabilizerness from the $|\mathbb{Z}_2\rangle$ initial state, in this case the non-stabilizerness is largely due to local basis rotations.

\subsection{The origin of non-stabilizerness}\label{Sec:unitary}

\begin{figure}[tb]
    \centering
    \includegraphics[width=0.99\linewidth]{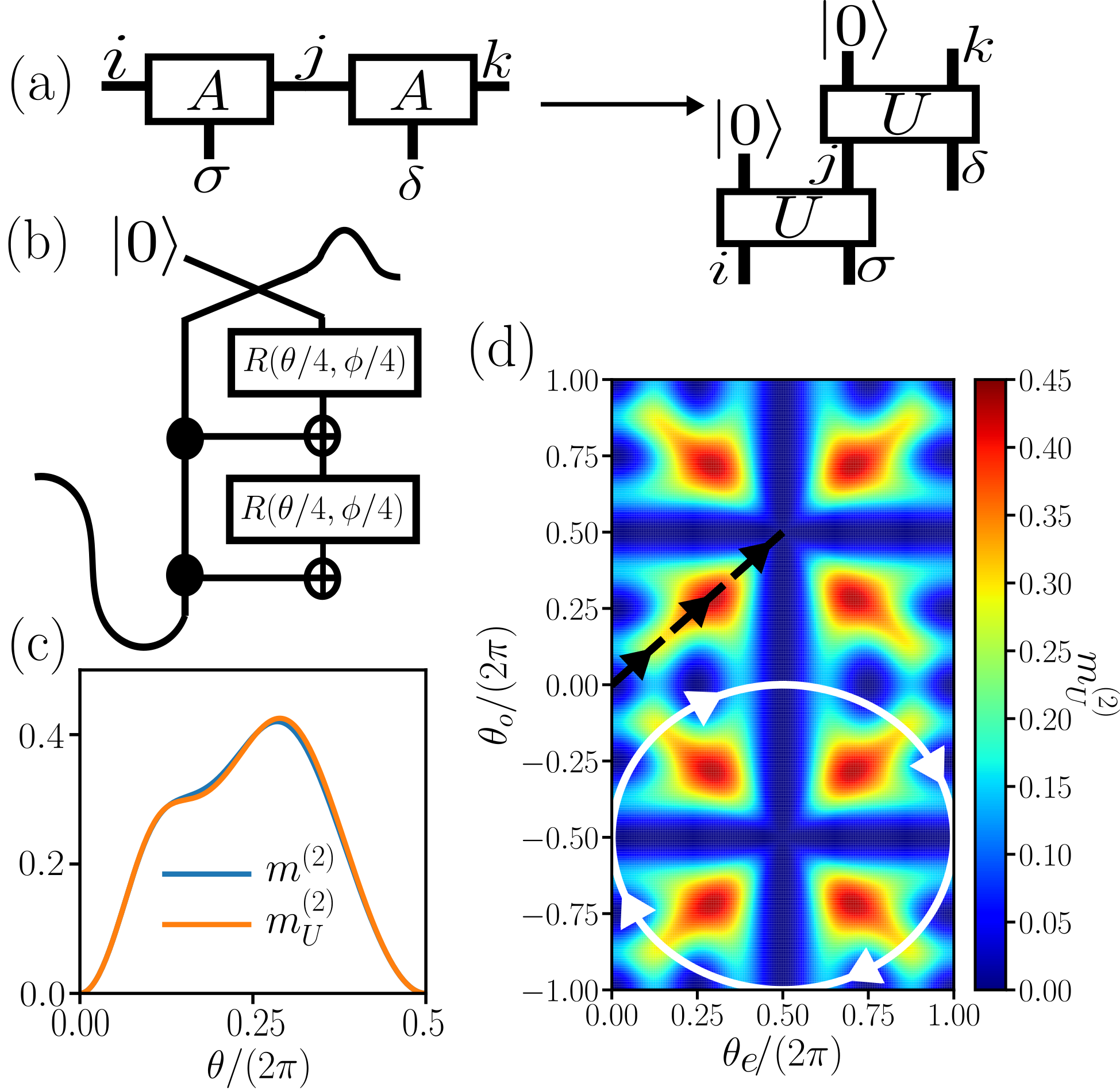}
    \caption{(a)-(b): Decomposing the MPS tensors into unitary rotations (a), with the resulting PXP unitary circuit shown in (b).
    (c): The SRE $m^{(2)}$ of the MPS ansatz along the $\ket{0}$ trajectory is well-described by $m^{(2)}_{U}$, Eq.~(\ref{eq:mU}), the non-stabilizerness of the input state $\rho_\mathrm{in}$ after being acted on by the unitary generated by the ansatz.  (d) The SRE from the PXP unitary ansatz, $m^{(2)}_{U}$, across the entire manifold $\mathcal{M}$. 
    }
    \label{fig:Unitary}
\end{figure}

To provide an intuitive understanding of large SRE in Fig.~\ref{fig:PXP Magic}(c), we utilize a quantum circuit description. Any MPS can be rewritten to satisfy the canonical condition~\cite{SCHOLLWOCK201196,Vidal2003}
\begin{equation}
 \sum_\sigma (A^\sigma)^\dagger A^\sigma=\mathbb{1}_{\chi},   
\end{equation}
which can be thought of as $\chi$ columns of a $d\chi \times d\chi$ unitary matrix. The product structure of the MPS wave function, $\cdots A^{\sigma_j}A^{\sigma_{j+1}}\cdots$, is reproduced by composing these unitary matrices in a staircase pattern as shown in Fig.~\ref{fig:Unitary}(a), with one leg contracted with a dummy state $\ket{0}$ to select the first $\chi$ columns of the unitary. The ansatz described in Eq.~\eqref{Eq:PXP MPS} is no exception -- it can be constructed from the two-qubit unitary given in Fig.~\ref{fig:Unitary}(b):
\begin{equation}\label{Eq: PXP Unitary}
U_\mathrm{MPS} = \text{CNOT} \; 
 R^{(2)}_{\theta/4,\phi/4}  \;   \text{CNOT} \; R^{(2)}_{\theta/4,\phi/4} \;  \text{SWAP},
\end{equation}
where $R^{(2)}_{\theta,\phi} = \exp(i(\pi/2) \boldsymbol{n} \cdot \boldsymbol{\sigma})$ is a rotation of the second qubit around the axis $\boldsymbol{n}=(\sin\theta\cos\phi, \sin\theta\sin\phi, \cos\theta)$. Note that the $\theta$ and $\phi$ parameters in Eq.~\eqref{Eq: PXP Unitary} are the same as in Eq.~\eqref{Eq:PXP MPS}. 

$U_\mathrm{MPS}$ in Eq.~(\ref{Eq: PXP Unitary}) is simply an arbitrary controlled rotation on the second qubit, conditional on the first being $\ket{0}$, and combined with a SWAP gate. The unitary is made up of two CNOT gates and one SWAP gate which are Clifford gates, and two local rotations which are non-Clifford and responsible for the non-stabilizerness of the MPS wave function. The subtlety, however, is that the non-Clifford unitaries are only relevant if they act non-trivially on the input state. The first qubit always acts upon $\ket{0}$ by construction. In the thermodynamic limit, the second qubit acts on $\rho_R$, the right eigenvalue of the MPS transfer matrix $E{=} \sum_{\sigma} A^\sigma {\otimes} \bar{A}^\sigma$. Thus, the non-stabilizerness of the MPS can be estimated as  the non-stabilizerness of $\rho_\mathrm{in}{=}\ket{0}\bra{0}\otimes \rho_R$ with $U_\mathrm{MPS}$ applied to it, minus the non-stabilizerness of $\rho_\mathrm{in}$,
\begin{equation}\label{eq:mU}
m^{(2)}_U=m^{(2)}(U_\mathrm{MPS} \rho_\mathrm{in} U^\dagger_\mathrm{MPS})-m^{(2)}(\rho_R),
\end{equation}
where we have taken advantage of the SRE additivity under tensor product to write $m^{(2)}(\rho_\mathrm{in})=m^{(2)}(\rho_R)$. Due to the inherent gauge freedom in our MPS, $U_\mathrm{MPS}$ is only unique up to a unitary transformation, $U_\mathrm{MPS}\rightarrow (u \otimes \mathbb{I}_d) U_\mathrm{MPS} (u \otimes \mathbb{I}_d)$, where $u$ is an arbitrary $\chi {\times} \chi$ unitary matrix. Since $u$ can be non-Clifford, it can change $m^{(2)}_U$. The gauge chosen throughout this paper is the only natural choice for the PXP model: the form of $\rho_\mathrm{in}$ and $U_\mathrm{MPS}$ guarantees that the $\ket{11}$ component of the MPS remains zero, ensuring that the Rydberg blockade constraint is also obeyed by the unitary generating the MPS.

In Fig.\ref{fig:Unitary}(c) we compare the non-stabilizerness obtained from the MPS ansatz in Fig.~\ref{fig:PXP Magic}(b)  against $m^{(2)}_{U}$ in Eq.~\eqref{eq:mU}.  We observe excellent agreement between the two over the range in which we traverse diagonally through one of the arrow-shaped regions, especially at the two extremes of the $\theta$ range. In the region of high non-stabilizerness, our ansatz in Eq.~\eqref{eq:mU} very slightly deviates from the exact $m^{(2)}$, implying that $U_\mathrm{MPS}$ is not the exact unitary and further corrections may be needed to exactly capture the SRE across the entire manifold $\mathcal{M}$. In Fig.~\ref{fig:Unitary}(d), $m^{(2)}_U$ is shown for the full two-site unit-cell ansatz. We see that, despite small deviations, it is significantly closer to the full SRE than 2-site SRE and accurately recreates the SRE of the PXP manifold. 

\begin{figure*}[bt]
    \centering
    \includegraphics[width=0.99\linewidth]{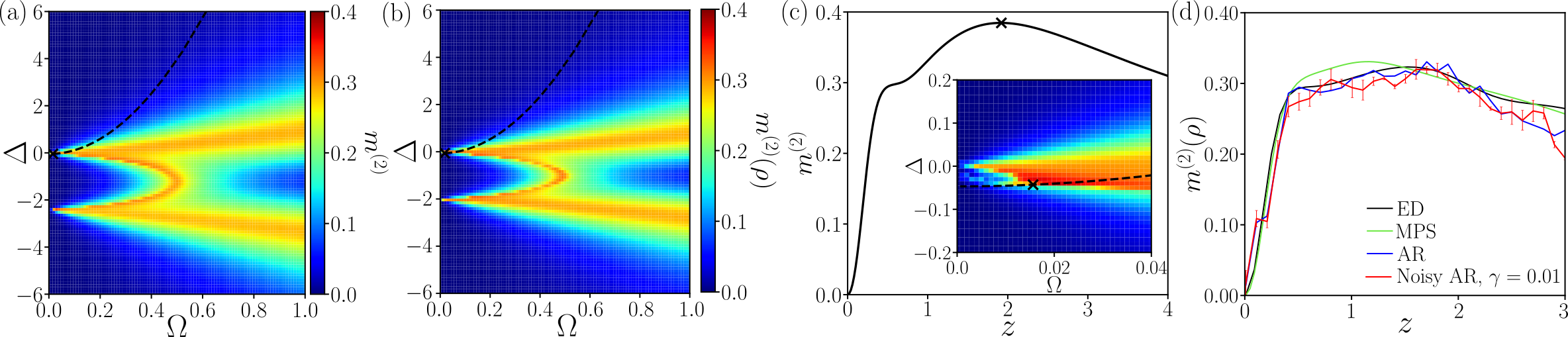}
    \caption{(a)-(b): SRE of the ground state of the Rydberg model in Eq.~(\ref{Eq: Rydberg Model}) with $V{=}1$ and varying $\Delta$ and $\Omega$, with  interactions truncated at next-nearest neighbors. The ground state of $N=51$ atoms was obtained using density matrix renormalization group (DMRG) based on ITensor Library~\cite{ITensor} at a bond dimension of $\chi=8$. Panel (a) shows the SRE $m^{(2)}$, while panel (b) is the two-site reduced density matrix SRE $m^{(2)}(\rho)$.
    Dashed line represents the trajectory of the effective model $H_0$ in Eq.~\eqref{Lesanovsky}, with the cross denoting the point of maximum non-stabilizerness. 
    (c): Non-stabilizerness in the effective model $H_0$, Eq.~(\ref{eq:parametrization}), as a function of $z$. The maximum value $m_2\approx0.38$ is comparable to Fig. \ref{fig:PXP Magic} and attained at $z\approx2$ (cross). Inset is a magnified section of the phase diagram in panel (a).   
    (d) Two-site reduced density matrix SRE $m^{(2)}(\rho)$ for the ground state of the Rydberg model computed by four different methods: MPS (green line), ED (black line), clean adiabatic ramping (AR) protocol with sampling in Bloqade (blue line), and Bloqade's stochastic noise simulation with noise amplitude $\gamma=0.01$ (red symbols). All simulations were performed for $N=11$ atoms while the Bloqade simulations were performed with a lattice spacing of 3.12$\mu m$ and $n_\mathrm{shots}=1000$. The error bars represent the statistical error in averaging the noisy results over 10 runs. 
    }
    \label{fig: experimental phase diagram}
\end{figure*}

\section{Experimental protocols}\label{Sec: Experiment}

Finally, in this section we outline two protocols for experimentally accessing the non-stabilizer states in Fig.~\ref{fig:PXP Magic}. The PXP model, Eq.~(\ref{Eq:PXP model}), arises as a limiting case of the more general Hamiltonian describing a 1D array of Rydberg atoms~\cite{Bernien2017}
\begin{equation}\label{Eq: Rydberg Model}
    H_\mathrm{Ryd} = \frac{\Omega}{2} \sum_{j} \sigma^{y}_j + \Delta \sum_j n_{j} + V \sum_{i<j} \frac{n_{i}n_{j}}{|i-j|^\alpha},
\end{equation}
with $n_{j}=|1\rangle\langle1|_j$ counts local excitations, $\Delta$ is the chemical potential, and $V$ is the overall strength of the van der Waals interactions. Due to the fast decay of the interactions ($\alpha=6$), we will neglect interaction terms beyond next-nearest neighbors. The Hamiltonian $H_\mathrm{Ryd}$ reduces to the PXP model in the regime $V \gg \Omega, \Delta$~\cite{FendleySachdev,LesanovskyKatsura,Turner2018}.

In the strong blockade regime $V \gg \Omega, \Delta$, the non-stabilizerness in Fig.~\ref{fig:PXP Magic} can be accessed by repeating the same type of quench experiments performed in Ref.~\cite{Bernien2017}.  While measuring the global SRE may not be feasible, the SRE can be approximated locally using a two-site reduced density matrix, Eq.~(\ref{Eq:mixed_state_SRE}), as we have demonstrated in Fig.~\ref{fig:PXP Magic}. Note that, in order for the ground state of  Eq.~\eqref{Eq: Rydberg Model} to be consistent with states obtained through dynamics in Fig.~\ref{fig:PXP Magic}(c), where $\langle \sum_j \sigma^x_j\rangle=0$, we have oriented the Rabi flip term in Eq.~\eqref{Eq: Rydberg Model} along the $y$-axis. This, however, does not affect the SRE value.

Alternatively, it is possible to access the non-stabilizerness in Fig.~\ref{fig:PXP Magic} by adiabatic ground state preparation. 
With the reparametrizations 
\begin{eqnarray}\label{eq:parametrization}
\Omega = 2V/(2^{\alpha}z), \quad \Delta = - (2V/2^{\alpha})(3-1/z^2),    
\end{eqnarray}
the ground state of $H_\mathrm{Ryd}$ can be approximated with the ground state of the following Hamiltonian~\cite{Lesanovsky2011,Lesanovsky2012}: 
\begin{eqnarray}\label{Lesanovsky}
    H_{0} =\sum_{j=1}^{N} P_{j-1}\left( \sigma^{y}_{j}+zP_{j}+z^{-1}n_{j}\right) P_{j+1}.
\end{eqnarray}
The advantage of $H_0$ is that it is frustration-free and its ground state is known exactly for any value of $z$~\cite{Lesanovsky2011}. In Appendix~\ref{subsec:parent hamiltonian}, we prove that the ground state of $H_0$ is, in fact,  equivalent to the MPS state in Eq.~\eqref{Eq:PXP MPS} with 
\begin{eqnarray}
\theta=\theta_o=\theta_e, \quad z=\sin(\theta/2)/\cos^2(\theta/2).    
\end{eqnarray}
Thus, the non-stabilizerness in Fig.~\ref{fig:PXP Magic} can be generated by simply preparing the ground state of $H_0$ with $z\approx2$.

In Fig.~\ref{fig: experimental phase diagram}(a) we show the SRE of the ground state of $H_\mathrm{Ryd}$ for $N=51$ atoms as a function of $\Omega$ and $\Delta$, with fixed $V=1$. We overlay the trajectory of $H_{0}$ by the dashed black line and mark the point of maximal non-stabilizerness with a cross. We observe a large W-shaped band of non-stabilizerness which becomes more pronounced at larger $\Omega$ with the trajectory touching the tip of this band. A similar phase diagram is observed in Fig.~\ref{fig: experimental phase diagram}(b), where we plot the two-site reduced density matrix SRE for the ground state of the Rydberg model. We find this local measure of magic produces an excellent qualitative agreement with the full calculation of the SRE in Fig.~\ref{fig: experimental phase diagram}(a), capturing the W-shaped band emanating from $\Delta\approx0$. 

The black dashed line in Fig.~\ref{fig: experimental phase diagram}(a),(b) represents the effective Hamiltonian for the MPS ansatz, $H_0$, whose trajectory is analyzed in more detail in  Fig.~\ref{fig: experimental phase diagram}(c). The SRE of the trajectory of $H_{0}$ attains a maximum value of $m^{(2)}\approx0.38$ at $z\approx2$, corresponding to $\Omega \approx 0.0156$ and $\Delta\approx-0.0429$. 
Similarly, the  maximal value of the two-site density matrix SRE, $m^{(2)}(\rho)\approx0.32$, in Fig.~\ref{fig: experimental phase diagram}(b) is comparable to the actual SRE maximum of $m^{(2)}\approx0.38$, suggesting that quantum state tomography would allow to accurately approximate the SRE.

Finally, in Fig.~\ref{fig: experimental phase diagram}(d) we  demonstrate that it is possible to detect non-stabilizerness by preparing the states along the trajectory described by the Hamiltonian $H_0$ and by measuring the mixed state magic of two sites at the center of the chain. To assess the impact of experimental imperfections on the measurements of non-stabilizerness, we perform realistic small-scale simulations of QuEra's Rydberg atom platform using the Bloqade interface~\cite{bloqade2023quera}.  We determine the ground state $|\psi\rangle$ of the Rydberg Hamiltonian along the desired trajectory using exact diagonalization (ED) and by adiabatically ramping (AR) from the polarized state $|0\rangle$. We have assumed a maximum ramp time of $4 \mu s$~\cite{bloqade2023quera}. To adiabatically prepare the ground state along the $H_0$ trajectory, we consider two seperate ramp sequences depending on the sign of $\Delta$. For $\Omega$ we consider a quick linear ramp to the desired $\Omega_{i}$. When we ramp to negative $\Delta$ we also consider a linear function where we ramp from $\Delta_\mathrm{prep}$ to the desired $\Delta_{i}$. When ramping to a positive $\Delta_i$ we consider a sigmoid ramp with curvature $k$
\begin{equation}
    \Delta(t)=\frac{t-kt}{k-2k|t|+1} \;,
\end{equation}
with curvature $k=0.98$. Secondly we shift the inflection point such that it lies at $(t,\Delta)=(2,0)$ and we rescale $t\in(0,2)$ and $t\in(2,4)$ to match $\Delta_{prep}$ and $\Delta_{i}$ respectively. To mimic the experimental protocol, we perform two-site tomography by sampling the probability distribution $|\psi|^2$ and calculating the 16 possible Pauli expectation values for two sites located at the center of the chain. Using these measurements, the definition given in Eq.~\eqref{Eq:mixed_state_SRE} and standard statistical error analysis, we obtain estimates of the $n=2$ mixed state SRE. To simulate experimental noise, we consider a depolarizing noise channel with error rate $\gamma$ during the adiabatic ramping procedure and optimize the ramp parameters over the trajectory to obtain the mixed state SRE. 

Figure~\ref{fig: experimental phase diagram}(d) compares the mixed-state non-stabilizerness $m^{(2)}(\rho)$ obtained using ED and adiabatic ramping through noiseless and noisy channels with sampling (averaged over 10 runs). We also show the MPS results for $\chi=100$ for comparison and the error bars represent the average error over the 10 runs. We note that the difference between the peak of non-stabilizerness in Fig.~\ref{fig: experimental phase diagram}(d) and $z\approx 2$ in Fig.~\ref{fig: experimental phase diagram}(c) is due to finite-size effects and divergences in regions of large global non-stabilizerness. We find good agreement between the expected ED results and the experimental simulations in the regions of high non-stabilizerness. This suggests that our protocol can be implemented in practice, even after accounting for experimental noise and readout errors.

\section{Conclusions}\label{Sec:Conclusions}

We showed that kinetic constraints stemming from the Rydberg blockade can lead to many-body non-stabilizerness. While we focused on the strong blockade regime described by the PXP model, our approach can be directly generalized to other blockade regimes explored in recent experiments~\cite{zhao2024observationquantumthermalizationrestricted}. While in the main text we studied dynamical manifestations of non-stabilizerness, further results on the non-stabilizerness of PXP eigenstates can be found in Appendix~\ref{Sec:Eigenstates}. Utilizing a quantum circuit construction, we provided an intuitive understanding of the origin of non-stabilizerness of an MPS in terms of the number of non-stabilizer unitaries used to generate it. 
While we expect this analysis to have broader applicability beyond the PXP model, to make it more rigorous, it would be necessary to better understand the impact of the MPS gauge degree of freedom on non-stabilizerness.  

Our work highlights the fact that some low bond-dimension MPS can exhibit high levels of non-stabilizerness. Since all 1D area-law entangled states -- including the ground-states of local, gapped Hamiltonians --  can be arbitrarily well approximated by MPS~\cite{Verstraete2006,Hastings2007}, they can be efficiently generated using classical algorithms such as DMRG. Most of these wave functions possess intrinsic non-stabilizerness, which implies that there may be certain ground states which are easy to generate classically but difficult to create in a fault-tolerant quantum computer, in particular the points of maximum SRE identified for the Rydberg model above. Finally, an intriguing question is which many-body systems saturate the upper bound of non-stabilizerness. In this context, Rydberg atom arrays exhibit a variety of exotic  phases~\cite{Keesling2019,Semeghini2021} and it would be interesting to study the non-stabilizerness in their ground states to see if any of them further approach the upper bound.

\begin{acknowledgments}

We would like to thank Abolfazl Bayat, Benjamin B\'eri, Jad Halimeh, Ana Hudomal, David Jennings, and Jie Ren for inspiring discussions.  We acknowledge support by the Leverhulme Trust Research Leadership Award RL-2019-015 and the EPSRC Grant EP/Z533634/1. Statement of compliance with EPSRC policy framework on research data: This publication is theoretical work that does not require supporting research data. This work made use of the High Performance Computing facilities at the University of Leeds, UK. ZP acknowledges support by grant NSF PHY-2309135 to the Kavli Institute for Theoretical Physics (KITP).

\end{acknowledgments}

\appendix

\section{Calculating SRE by Pauli basis conversion}\label{App:pauli basis}

Calculating the SRE for an MPS state by the replica trick becomes infeasible for states with large bond dimensions due to the replicated state scaling as $\chi^{2n}$. Instead, this method can be reformulated to allow truncation of the bond dimension during the calculation.

For a normalized state $|\psi\rangle$ which is represented by an MPS of physical dimension $d$ and bond dimension $\chi$, we can define its density matrix as $\rho=|\psi\rangle\langle\psi|$, an MPO of bond dimension $\chi^{2}$ and two physical dimensions $d$ and $d^{\prime}$. We then introduce a three-legged Pauli tensor $P^{\alpha}_{i,i^{\prime}} = \frac{1}{\sqrt{2}}(\sigma^{\alpha})_{i,i^{\prime}}$ with bond dimension $\chi=d$ and physical dimension $d^{2}$. This Pauli tensor is then contracted on both physical bonds on each site of $\rho$, thereby converting it into the Pauli basis and leaving an MPS $|P(\psi)\rangle$ with bond dimension $\chi^{2}$ and physical dimension $d^{2}$. This now grants us the ability to truncate the bond dimensions to some $\chi_{P}$ if required.

Now that the MPS has been converted to the Pauli basis, we can perform the replica trick at a reduced cost. To do this we define a diagonal operator $W$ whose diagonal elements are just the components of the Pauli MPS at each site. This MPO is built by contracting a three-legged delta tensor of dimensions $d^2$. This delta tensor is contracted onto each physical leg of the Pauli MPS to produce the MPO $W$. For some stabilizer Renyi index $n$ we apply the MPO $W$, $n-1$ times onto $|P(\psi)\rangle$ to obtain $|P^{(n)}(\psi)\rangle = W^{n-1}|P(\psi)\rangle$, while also truncating the bond dimension after each application of $W$. Therefore the $n$th order SRE is given by:
\begin{equation}
    M^{(n)} = \frac{1}{1-n}\log(\langle P^{(n)}(\psi)|P^{(n)}(\psi)\rangle) - N,
\end{equation}
and, similarly, the SRE density is $m^{(n)} = M^{(n)}/N$.

\section{Parent Hamiltonian for MPS ansatz}\label{subsec:parent hamiltonian}

Any matrix-product state $\ket{\psi(A)}$ is the ground state of a so-called parent Hamiltonian which is a local, frustration-free Hamiltonian constructed as the sum of projectors 
\begin{eqnarray}
H^\mathrm{parent}=\sum_j \mathbb{P}_j, \quad \mathbb{P}_j^2=\mathbb{P}_j.    
\end{eqnarray}
The MPS of interest is in the null-space of the projector $\mathbb{P}_j$ and therefore has energy exactly $E=0$ with respect to this parent Hamiltonian. We would like to find a parent Hamiltonian for the MPS used in Eq.~(\ref{Eq:PXP MPS}) the main text.  In this section, we will work in the thermodynamic limit and assume a single-site unit cell, as we are interested in the dynamics of $\ket{0}$ state.

The projector which defines the parent Hamiltonian for the MPS in Eq.~\eqref{Eq:PXP MPS} is found by constructing the reduced density $\rho_n$ of $\ket{\psi(A)}$ over $n$ sites. The value of $n$ should be chosen such that $\rho_n$ has at least one eigenvalue zero. This is typically the case when $d^n > \chi^2$. For the PXP model, the Hilbert space obeys the Rydberg blockade constraint which excludes certain states, hence $\rho_n$ does not grow exactly like $d^n$. Nevertheless, $n=3$ is suitable since $\rho_n$ has 5 eigenvalues, which is larger than the $\chi^2=4$. 

We choose the basis $\{\ket{000},\ket{001},\ket{010},\ket{100},\ket{101}\}$ for $\rho_3$. The leading left/right eigenvalues of the MPS transfer matrix are 
\begin{eqnarray}
(\mathbb{L}|=(1,0,0,1), \quad |\mathbb{R})=\frac{1}{1+s^2}(1,cs,cs,s^2),  
\end{eqnarray}
where we have introduced a shorthand notation $c\equiv \cos(\theta/2)$ and $s\equiv \sin(\theta/2)$. Furthermore, $\rho_3$ can be shown to be:
\begin{widetext}
    \begin{equation}
    \rho_3=\frac{1}{1+s^2}\begin{pmatrix}
    c^4 & -i c^4 s e^{i\phi} & -i c^2 s e^{i\phi} & -i c^4 s e^{i\phi}& -c^4 s^2 e^{2i\phi} \\
    ic^4se^{-i\phi} & c^2 s^2 & c^2 s^2 & c^4 s^2 &-i c^2 s^3 e^{i\phi} \\
    ic^2se^{-i\phi} & c^2 s^2 & s^2 & c^2 s^2 & -i c^2 s^3 e^{i\phi}
    \\
    ic^4se^{-i\phi} & c^4 s^2 & c^2 s^2 & c^2 s^2 & -i c^2 s^3 e^{i\phi}\\
    -c^4 s^2 e^{2i\phi} & i c^2 s^3 e^{-i\phi} & i c^2 s^3 e^{-i\phi} & i c^2 s^3 e^{-i\phi} & s^4 \\
    \end{pmatrix}.
    \end{equation}
\end{widetext}
Thus, $\rho_3$ has a single eigenvalue zero with eigenvector:
\begin{equation}
\ket{\phi}= \frac{1}{\sqrt{1+s^2/c^4}} (
ie^{i\phi} s/c^2, 
0,
1,
0, 
0)^\mathrm{T}.
\end{equation}
If we define $z = s/c^2$, then the parent Hamiltonian of the MPS ansatz is $\mathbb{P}_j=\ket{\phi}\bra{\phi}$, where
\begin{eqnarray}
\nonumber    \mathbb{P}_j &=& \frac{1}{1+z^2}\Big(z^2\ket{000}\bra{000}+\ket{010}\bra{010} \\
\nonumber    &-& ize^{-i\phi}\ket{000}\bra{010}
    + ize^{i\phi}\ket{010}\bra{000}\Big)_{j-1,j,j+1}\\
\nonumber    &=& \frac{z}{1+z^2}P_{j-1} \Big( z P_j + z^{-1} n_j + \cos(\phi)\sigma^y_j  \\
 &+& \sin(\phi)\sigma^x_j \Big) P_{j+1}.
\end{eqnarray}
When $\phi=0$, this is the model introduced in Ref.~\cite{Lesanovsky2011} and discussed in the main text, up to an overall prefactor. Although considerably more tedious, it is possible to generalize this to the case of 2-site unit cell which would be needed for describing the dynamics of the $\ket{\mathbb{Z}_2}$ state. We have not pursued this, however, since the non-stabilizerness along the $\ket{\mathbb{Z}_2}$ orbit is smaller than for the $\ket{0}$ state.

\section{Robustness of Magic}\label{Sec:ROM}

In Sec~\ref{sec: SRE} we introduced the mixed state SRE for a two-site density matrix as an experimental "witness" for calculating SREs. However, there are other potential magic monotones that could be considered. In addition to the mixed state SRE, we also studied a different measure for local non-stabilizerness called the ``Robustness of Magic" (RoM)~\cite{Howard2017,Ahmadi2018,Sarkar_2020} which quantifies the minimal weight of stabilizer states that yields a stabilizer state when mixed with a density matrix $\rho$. Since the set of stabilizer states is overcomplete, it is possible to decompose $\rho$ into a weighted sum of stabilizer states, $\rho=\sum_{i}X_{i}S_{i}$, where $S_{i}$ is a stabilizer state, while satisfying the normalization constraint $\sum_iX_i=1$. For $\rho$ to have non-stabilizerness, at least one of the weights must be negative. Therefore the ROM is expressed as a convex optimization problem:
\begin{equation}
    \mathcal{R} = \underset{\{X_i\}}{\text{inf}}\{\sum_i |X_i|-1:AX=B\},
\end{equation}
where $A_{\alpha\beta} = \mathrm{tr}(\sigma^{\alpha}S_{\beta})$  and  $B_{\alpha} = \mathrm{tr}(\sigma^{\alpha}\rho)$, with $\sigma^{\alpha}$ as a \emph{two-qubit} Pauli string. However, the RoM is known to be multiplicative under tensor product, therefore we instead define the ``log-free'' RoM $\ln (\mathcal{R}+1)$~\cite{Liu2022} which is additive under tensor product  and allows for a better comparison with SREs.

\begin{figure}[tb]
    \centering
    \includegraphics[width=0.99\linewidth]{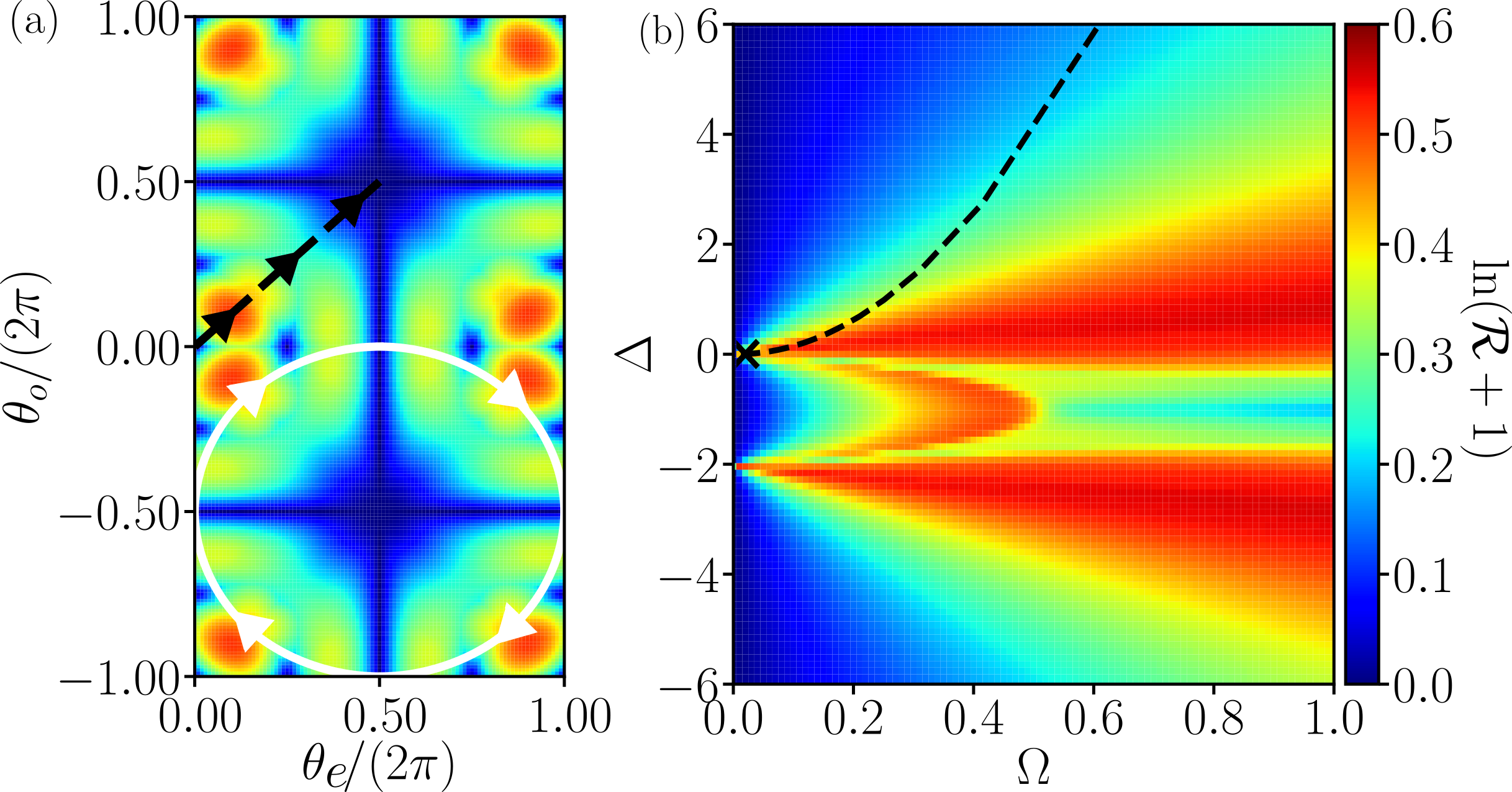}
    \caption{(a): Phase diagram for the log-free RoM $\ln(\mathcal{R}+1)$ across the MPS manifold $\mathcal{M}$. (b): Phase diagram of the log-free RoM for the ground state of the Rydberg model in Eq.~(\ref{Eq: Rydberg Model}). Simulations were performed for $N=51$ atoms at $\chi=8$ using iTensor~\cite{ITensor} with fixed $V=1$ and interactions truncated beyond next-nearest neighbors. 
    }
    \label{fig:robustness of magic}
\end{figure}

Figure~\ref{fig:robustness of magic}(a) shows the RoM for the two-site unit cell MPS in Eq.~(\ref{Eq:PXP MPS}). While the RoM captures some of the arrowhead-like structure of the SRE, it diverges dramatically from the SRE in the long-range non-stabilizerness regions we identified in Fig.~\ref{fig:PXP Magic}(d). Noticeably, a large area of stabilizerness is created at the centre of the unit cell where the points of the arrow-heads should be located. In Fig.~\ref{fig:robustness of magic}(b) we plot the phase diagram of the log free robustness of magic, $\ln(\mathcal{R}+1)$, for the ground state of the Rydberg model. This measure produces more qualitative accuracy with the original phase diagram in Fig.~\ref{fig: experimental phase diagram}(a) compared to the two site PXP phase diagram in Fig. \ref{fig:PXP Magic}(c). Therefore, the RoM could also be used as a measure of local non-stabilizerness similar to mixed state SRE's. This metric is able to approximately capture the W-shaped band of non-stabilizerness that emanates from $\Delta\approx0$ and extends to larger $\Omega$.

\section{Non-stabilizerness of PXP eigenstates} \label{Sec:Eigenstates}

In the main text we focused on the PXP model and explored the non-stabilizerness generated under its quantum dynamics. Here we turn our attention to the non-stabilizerness of energy eigenstates of the PXP model. One of the remarkable properties of this model is that it has a few eigenstates in the middle of the energy spectrum that can be written as exact low bond-dimension MPS~\cite{Lin2019}. This will allow us to directly evaluate their SRE with the replica MPS approach outlined above and in the main text.

First, we analyze two exact scarred eigenstates with energy $E=0$ exactly in the middle of the spectrum.
Assuming even chain lengths and periodic boundary conditions (PBCs), these states were written down in Ref.~\cite{Lin2019} using the following MPS ansatze:
\begin{eqnarray}\label{eq:PXPMPSE0}
|\psi_{1}\rangle &=& \sum_{\{\sigma_i\}} \mathrm{tr}(B_{1}^{\sigma_{1}}C_{2}^{\sigma_{2}}...B_{1}^{\sigma_{N-1}}C_{N}^{\sigma_{L}})|\sigma_{1}\ldots\sigma_{N}\rangle,
\\
|\psi_{2}\rangle  &=& \sum_{\{\sigma_i\}} \mathrm{tr}(C_{1}^{\sigma_{1}}B_{2}^{\sigma_{2}}...C_{1}^{\sigma_{N-1}}B_{N}^{\sigma_{N}})|\sigma_{1}\ldots\sigma_{N}\rangle,
\end{eqnarray}
with the MPS matrices given by
\begin{eqnarray}\label{eq:PXPMPSB}
   B^{0} &=& \begin{pmatrix}
        1&0&0\\
        0&1&0
    \end{pmatrix}, \;\; 
    B^{1}=\sqrt{2}\begin{pmatrix}
        0&0&0\\
        1&0&1
    \end{pmatrix}, \\ \label{eq:PXPMPSC}
    C^{0}&=&\begin{pmatrix}
        0&-1\\
        1&0\\
        0&0
    \end{pmatrix}, \;\;
    C^{1}=\sqrt{2}\begin{pmatrix}
    1&0\\
    0&0\\
    -1&0
    \end{pmatrix}.
\end{eqnarray}
For $|\psi_{1}\rangle$ and $|\psi_{2}\rangle$, we find that $m^{(2)}\approx0.376$, which is significantly larger than, e.g., the SRE at the critical point of the Ising model. However, we find that the SRE of these eigenstates is slightly smaller than maximal SRE we found in the main text.

\begin{figure}[tb]
    \centering
    \includegraphics[width=0.99\linewidth]{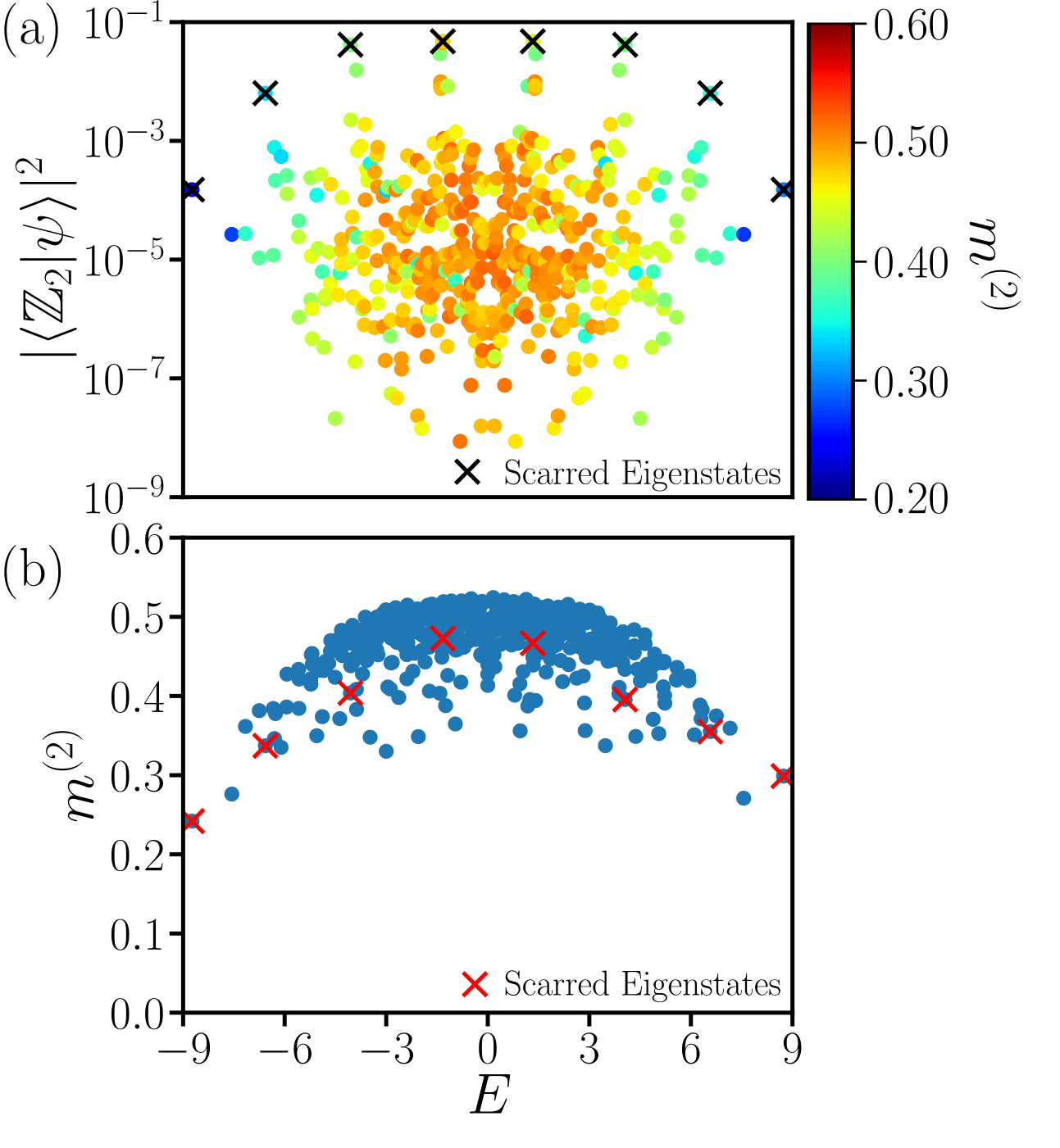}
    \caption{(a) Overlap between the $\ket{\mathbb{Z}_{2}}$ and all eigenstates of the PXP model plotted as a function of their energy. Each eigenstate is colored by its $m^{(2)}$ value (color bar) obtained by Monte Carlo sampling. (b) SRE of the PXP eigenstates as a function of their energy $E$. In both panels, the crosses indicate the scarred eigenstates. The simulations were performed for the PXP model with $N=14$ spins with OBCs and in the $P=+1$ parity sector, using 300000 Monte Carlo samples. 
    }  
    \label{fig:PXP eigenstates}
\end{figure}

Apart from the MPS states in Eqs.~\eqref{eq:PXPMPSE0}-\eqref{eq:PXPMPSC}, the majority of PXP eigenstates are believed to be volume-law entangled~\cite{TurnerAPS}. Hence, we will rely on Markov chain Monte Carlo sampling to evaluate their SRE. This method overcomes the $4^{N}$ scaling of the exact evaluation of the SRE, but it comes at the cost of generally poor convergence with the number of samples required for accurate convergence being $N_{s}\approx 1\times10^{6}$. 

At energy $E=0$ in the middle of the spectrum, similar to the MPS states in Eqs.~\eqref{eq:PXPMPSE0}-\eqref{eq:PXPMPSC} above, there are also some volume-law entangled states that can be written down in analytic form. For example,  for a spin-$1/2$ PXP chain with PBCs and size $N=2L$, where $L$ is the size of the half chain, an exact volume-law entangled eigenstate at energy $E=0$ is given by the following ``rainbow''-like state~\cite{ivanov2024}:
\begin{equation}\label{Eq:Volume state}
    \ket{E}=\frac{1}{\sqrt{|F_L|}}\sum_{f\in F_{L}}(-1)^{|f|}\ket{f}_{1,...,L}\otimes\ket{f}_{L+1,...,2L}
\end{equation}
where $F_{L}$ is the set of bitstrings for a chain of $L$ spins with PBCs and respecting the Rydberg blockade constraint. The normalization $|F_{L}|=\varphi_{L-1}+\varphi_{L+1}$ is given in terms of Fibonacci numbers, $\varphi_{n}$, while $|f|$ denotes the parity of a given bitstring. Performing the Monte Carlo sampling, we computed the SRE of the $\ket{E}$ state for several system sizes $L$, obtaining the $L\to\infty$ extrapolated SRE of  
$m^{(2)}=0.180\pm0.008$. This implies that, in the thermodynamic limit, the $\ket{E}$ state is closer to a stabilizer state compared, e.g., to the peak in the SRE of the critical Ising model. Intuitively, this may be expected due to the sparse  structure of the $\ket{E}$ state when we consider the Hilbert space of the entire $2L$ chain. Additionally, the non-zero amplitudes on basis states in $\ket{E}$ are uniform, requiring fewer non-Clifford gates to construct the state.

Finally, in Fig. \ref{fig:PXP eigenstates} we study the SRE of all eigenstates in the PXP model with $N=14$ spins. We use exact diagonalization (ED) to extract the eigenstates and then apply Monte Carlo sampling to obtain their SRE. Here we assume open boundary conditions (OBCs) while also resolving parity symmetry. We found that OBCs result in a much better convergence of the Monte Carlo sampler compared to PBCs. Fig. \ref{fig:PXP eigenstates}(a) shows the overlaps of eigenstates with the $\ket{\mathbb{Z}_2}$ state as a function of their energy $E$, with the color bar showing $m^{(2)}$ of each eigenstate. We also highlight the scarred eigenstates using crosses. The bulk of the spectrum contains a large density of thermal eigenstates with large non-stabilizerness. Towards the edges of the spectrum, we see a noticeable drop in $m^{(2)}$, indicating that the ground state and low-energy excitations are closer to stabilizer states.

To isolate the non-stabilizerness of the scarred states, we plot the SRE of the eigenstate spectrum in Fig.~\ref{fig:PXP eigenstates}(b) and also label the scarred states. The ground state of the spectrum has the lowest SRE of the spectrum with a value comparable to the SRE of the 1D critical Ising model~\cite{Tarabunga2023,Liu2024}. The scarred eigenstates exhibit a roughly linear increase of SRE until we reach the mid-spectrum eigenstates with the largest overlap on $\ket{\mathbb{Z}_2}$ state. We note that any asymmetry in the plots in Fig.~\ref{fig:PXP eigenstates} around $E=0$ is due to accidental degeneracies in the spectrum or convergence issues in Monte Carlo sampling.

\bibliography{references}

\end{document}